%% file: main.tex
\documentclass[runningheads]{llncs}
\usepackage[T1]{fontenc}
\usepackage{cite}
\usepackage{amsmath,amssymb,amsfonts}
\usepackage{algorithmic}
\usepackage{graphicx}
\usepackage{textcomp}
\usepackage{xcolor}
\usepackage{booktabs}
\usepackage[nolist]{acronym}
\usepackage[linesnumbered,ruled,boxed,commentsnumbered,noend]{algorithm2e}
\usepackage{graphicx}
\begin{acronym}[TDMA]
    \acro{ikd-Tree}{Incremental K-d Tree}
    \acro{kNN}{k-Nearest Neighbors}
    \acro{SWINN}{Sliding Window-based Incremental Nearest Neighbors}
    \acro{ANN}{Approximate Neighborhood Search}
    \acro{DES}{Dynamic Ensemble Selection}
    \acro{ML}{Machine Learning}
    \acro{SAM-kNN}{Self Adjusting Memory model for the kNN}
    \acro{DDCS}{Double Dynamic Classifier Selection}
    \acro{AB-DES}{Adaptive Bagging-based Dynamic Ensemble Selection}
    \acro{MOA}{Massive Online Analysis}
    \acro{LSH}{Local Sensitive Hashing}
    \acro{KNNENS}{kNN Ensemble-based Method}
    \acro{ML-SAM-kNN}{Multi-label kNN with Self Adjusting Memory}
\end{acronym}

\begin{document}
\title{Online K-d tree for approximate neighborhood search in data streams}
\author{Eduardo V. L. Barboza
% \orcidID{0009-0006-5556-2619} 
\and
Robert Sabourin
% \orcidID{0000-0002-9098-1011} 
\and
Rafael M. O. Cruz
% \orcidID{0000-0001-9446-1040}
}
% \author{Anonymous Authors}

%
\authorrunning{Barboza et al.}
% \authorrunning{}
\institute{LIVIA, École de Technologie Supérieure, Montréal QC, CA \\
\email{eduardo.lima-barboza.1@ens.etsmtl.ca}\\
\email{\{robert.sabourin, rafael.menelau-cruz\}@etsmtl.ca}}
% \institute{Anonymous Institute}
%
\maketitle              % typeset the header of the contribution
\begin{abstract}
The \ac{kNN} algorithm has long been widely used in \ac{ML} applications. However, the main concern when using it is the computational cost required for neighborhood search, which can make it unfeasible for large-scale applications. Optimization algorithms, such as the K-d tree, become an option in such scenarios. Under data streams, it can be challenging to maintain the properties of the K-d tree, as it requires inserting and deleting nodes on the fly. These operations can make maintaining the tree's balance and invariants difficult. Additionally, traditional K-d trees were initially designed for Minkowski-based distance functions. In this work, we describe an Online K-d tree and its adaptation to the Canberra distance that supports dynamic updates over data streams while preserving the structural invariants required for efficient traversal. Experimental analysis demonstrates that the Online K-d tree algorithm achieves faster processing time under data streams, and that adapting to the Canberra distance enabled effective subtree pruning, as evidenced by a minor loss in average accuracy and a substantial gain in instances processed per second. Our implementation can be found in our GitHub repository\footnote{https://github.com/eduardovlb/OKDTree}.
\end{abstract}
\input{introduction}
\input{relWork}
\input{proposal}
\input{experiments}
\input{conclusion}
\bibliographystyle{splncs04}

% \bibliography{references}

\end{document}

%% file: introduction.tex
\section{Introduction}

Neighborhood search is a fundamental component of many \ac{ML} methods, including clustering, oversampling, and dynamic ensemble selection \cite{chawla2002, ahmed2020, barboza2025}. Among these, the \ac{kNN} algorithm remains widely used due to its simplicity and effectiveness. However, its computational cost has motivated the development of optimization techniques that reduce the number of distance calculations by pruning the search space \cite{bentley1975}.

Most \ac{kNN} optimization methods were designed for static settings, where the full training set is available \textit{a priori}. In data streams, where instances arrive continuously, and models must support frequent insertions and deletions, these assumptions no longer hold \cite{Lukats2025}. Maintaining efficient neighborhood search under such dynamic conditions is particularly challenging, as structural invariants required for pruning may be violated over time.

The K-d tree \cite{bentley1975} is a classical space-partitioning structure that enables efficient neighborhood search by recursively splitting the feature space and pruning subtrees that cannot contain nearest neighbors. While approximate variants further reduce processing time by relaxing pruning criteria \cite{ram2019}, most existing approaches assume static data and provide limited analysis under streaming scenarios.

In this work, we describe adaptations to the K-d tree for data stream processing, namely the online K-d tree, and evaluate its behavior under dynamic updates and concept drift. The neighborhood search procedure is extended to support the Canberra distance, demonstrating that effective pruning can be maintained beyond standard Minkowski-based metrics. Nevertheless, as with other space-partitioning structures, this approach remains affected by the curse of dimensionality, and its benefits may diminish as dimensionality and data distribution characteristics become less favorable.

% This paper addresses the following research questions:
% \begin{itemize}
%     \item \textbf{RQ1:} How does the K-d tree behave under different data stream characteristics?
%     \item \textbf{RQ2:} How do insertion and deletion operations impact the K-d tree in streaming settings?
%     \item \textbf{RQ3:} What are the main limitations of using K-d trees for neighborhood search in data streams?
% \end{itemize}

This study builds on prior experimental observations of incremental dynamic ensemble selection in data streams \cite{barboza2025}, focusing on the role of approximate neighborhood search structures and distance functions in streaming classification. Our experiments show that the online K-d tree achieves substantially faster neighborhood computation than brute-force \ac{kNN} on average, with minimal loss in classification performance, making it a viable alternative for streaming applications.

%% file: relWork.tex
\section{Related Work}
\label{sec:related-work}

\ac{kNN} optimization methods typically aim to reduce the number of distance computations. 
The K-d tree \cite{friedman1977, bentley1975}, the focus of this work, organizes instances through recursive axis-aligned splits, enabling pruning of large regions of the search space. 
However, the original formulation assumes static datasets, and dynamic insertions or deletions may lead to unbalanced structures or violated invariants.

Alternative tree-based structures for neighborhood search include Ball Trees \cite{omohundro1989}, Cover Trees \cite{beygelzimer2006}, M-Trees \cite{ciaccia1997}, Vp-Trees \cite{yianilos1993}, and R-Trees \cite{guttman1984}. 
These approaches have been primarily studied in static settings and often require costly rebalancing or complex deletion mechanisms, limiting their suitability for high-throughput data streams. 
As a result, several works have focused on adapting K-d trees to dynamic scenarios.

Progressive K-d Trees \cite{jo2017} rebuild the structure based on quality degradation, while the \ac{ikd-Tree} \cite{cai2021} supports incremental updates through lazy deletion and selective subtree rebuilding. 
Parallel variants have also been proposed, including approaches supporting batch updates via parallel construction \cite{Men2025}. 
These methods, however, typically assume batched insertions and emphasize computational scalability rather than approximate neighborhood search under continuous streaming updates and sliding windows, which is the focus of this work. 
Beyond tree-based methods, alternative \ac{ANN} strategies include \ac{LSH} \cite{jafari2021} and graph-based \ac{kNN} approaches \cite{malkov2020, MASTELINI2024101979}.

In data stream settings, \ac{kNN}-based classifiers such as \ac{SAM-kNN} \cite{losing2016} and \ac{ML-SAM-kNN} \cite{roseberry2018} address concept drift through short- and long-term memory mechanisms. 
These methods primarily focus on drift adaptation rather than accelerating neighbor search and typically rely on exact distance computations within sliding windows. 
K-d trees have also been explored for stream clustering \cite{Erdin2024, XU20242557}, exploiting spatial locality to identify regions with homogeneous class distributions, suggesting that class-discriminative information may be spatially localized.

%% file: proposal.tex
\section{Online K-d Tree}
\label{sec:proposal}

Building K-d trees for data streams introduces challenges related to incremental updates, deletions, and evolving data distributions. Moreover, classic K-d trees assume Minkowski-based distance functions, necessitating adaptations when alternative metrics are more appropriate. In this work, we describe adaptations to the K-d tree to operate on data streams and support the Canberra distance, extending prior experimental observations reported in \cite{barboza2025}. The motivation for adapting the K-d tree to the Canberra distance is that it has been shown to deliver the best results under data streams, as reported in \cite{barboza2023}. While other distance functions may also be accommodated, their split criteria must be adjusted.
% accordingly.

% The idea of building a K-d tree algorithm that operates on data streams arose from various challenges in this scenario. First, the continuous arrival and insertion of data into the K-d tree can cause it to become unbalanced as the data distribution changes over time. In addition, deletion operations, which are crucial under data streams, can cause some issues with the structure of a K-d tree. Further, the classic K-d tree neighborhood search algorithm was designed to Minkowski-based distances. When one wants to use another distance function that may be more proper in a specific scenario, the K-d tree needs to be adapted. The Canberra distance, for example, has already been shown to deliver the best results in various data stream scenarios \cite{barboza2023}. In this work, we describe the adaptation of the K-d tree algorithm to work under data streams and to support the Canberra Distance function, extending the work from \cite{barboza2025}. Adaptations to other distance functions may be feasible, too, but the split criterion should be modified.

Now, let us define the invariants of the nodes in a K-d tree. Let $t$ be a node containing an instance $I_t = [\mathbf{x},y]$, a split dimension $s$, $t_l$ and $t_r$ being the children assigned to the left and right children of the node $t$, respectively. The invariants are as follows:

\begin{itemize}
    \item $\forall I \in t_l, I[s] < I_t[s]$
    \item $\forall I \in t_r, I[s] \geq I_t[s]$
\end{itemize}

% This implies that all of the instances in the left subtree of $t$ must have their values in the dimension $s$ smaller than $I_t[s]$. Otherwise, those children must be in the right subtree. The K-d tree data structure uses these invariants to organize the nodes into a binary structure, speeding up search operations.

% It uses this in neighborhood search to prune subtrees if its hyper-rectangle will not lead to a nearest neighbor. For approximate neighborhood search, the difference to the current split dimension compared to the current distance may be used.

% \subsection{Building the K-d Tree}

We construct the K-d tree using the standard recursive median-splitting procedure \cite{bentley1975}, incrementing the split dimension cyclically at each level. While prior work proposes heuristics such as selecting the dimension with the highest variance \cite{friedman1977, muja2009}, we avoid such heuristics because, in data streams, feature variance may change over time.

% The K-d tree in this work was built as follows: given a set of instances $S$ of size $n$, first sort the instances by the first dimension and compute the median. The first instance that carries the median becomes the root node. Instances with values smaller than the root node in the first dimension are assigned to the root's left subtree, and instances with values equal to or higher than the root node are assigned to the right. This process is repeated recursively at subsequent levels, incrementing the splitting dimension by 1 at each level. The overall construction of the K-d tree has a time complexity of $O(n\log{n})$.

% Other works also adopt heuristics to choose the dimension for the split dimension $s$, such as the dimension with the higher variance \cite{friedman1977, muja2009}. However, in this work, we do not adopt any heuristic. Instead, we increment the dimension by 1 at each level, since computing the variance across all instances can become computationally expensive when the number of instances is high, and the dimensionality is high. In data streams, faster building is preferable because we process instances on the fly, and the dimension with the highest variance can change over time.

% \subsection{Dynamic Operations and Approximate Neighborhood Search}

\noindent\textbf{Dynamic Operations.} Dynamic operations (insertion and deletion) are crucial for methods operating on data streams that rely on a sliding window. In the online K-d tree, the insertion of a node is performed as in the classic K-d tree algorithm \cite{bentley1975}. A new instance added to the structure traverses the levels of the K-d tree by comparing its feature values to the keys of the nodes in their respective split dimensions $s$. On a balanced K-d tree, this process has a time complexity of $O(\log{n})$.

The deletion of an instance in our online K-d tree is done by using a lazy deletion strategy. Instead of directly removing a node from the structure, it is flagged as deleted, preserving the structure of the K-d tree without further complexity. For a balanced K-d tree, this process also has a time complexity of $O(\log{n})$, equivalent to searching for a node. This strategy was employed before by \cite{cai2021}.

Given the frequent insertion of data samples, the organization of the nodes in the structure can become unbalanced with the arrival of new instances, thereby making the search for an instance and its neighbors less efficient. Furthermore, lazy deletion incurs memory and traversal overhead due to the accumulation of inactive (flagged) nodes within the structure. To overcome these issues, we adopt two criteria for rebuilding the K-d tree:

\begin{enumerate}
    \item When the tree is $\gamma$ times bigger than when it was constructed, rebuild it;
    \item When the proportion of deleted nodes exceeds a threshold $\beta$, rebuild the K-d tree only with the non-deleted nodes.
\end{enumerate}

These two strategies ensure that the K-d tree remains balanced while effectively managing memory usage.

% By employing the lazy deletion, the node $(7,4)$ would simply be flagged as deleted, and the structure would remain unchanged, avoiding the additional complexity to ensure consistency with respect to the invariants after node replacement, such as the reinsertion of instances. In data stream scenarios, where fast operations are crucial, lazy deletion is advantageous because it simplifies deletion in the K-d tree. In subsequent search operations, nodes flagged as deleted are ignored (distances are not calculated to them).

% Regarding the adopted deletion strategy, memory "garbage" can be created with the deletion of nodes. To overcome the above issues, we have added two different requisites to rebuild the K-d tree: 1) If the K-d tree is twice as big as when it was constructed, rebuild it from scratch, which is the same strategy adopted by \cite{muja2009}; 2) If the deleted nodes in the K-d tree are over a threshold $\beta$, the K-d tree is rebuilt from scratch with the non-deleted instances. These two strategies will deal with the memory garbage while maintaining a balanced K-d tree by rebuilding it when needed.

% \subsection{Approximate Neighborhood Search with K-d Tree}

\noindent\textbf{Approximate Neighborhood Search with K-d Tree.} The \ac{ANN} procedure used for the K-d tree is described in Algorithm~\ref{alg:getNeighbors}. 
At each visited node, the distance between the query instance $\mathbf{q}$ and the node instance is computed only if the node is active. The neighbors list is updated by inserting the instance when fewer than $k$ neighbors are available, or by replacing the farthest neighbor if a closer instance is found.

The search follows the subtree determined by the split dimension $s$, comparing $\mathbf{q}[s]$ to the node value. Backtracking is performed only when the split criterion is satisfied, allowing alternative subtrees to be explored. For Minkowski-based distances (Euclidean and Manhattan), the split criterion is defined as $c = |\mathbf{q}[s] - \mathbf{I_t}[s]|$, enabling pruning of subtrees. The source code is available in our GitHub repository.

% The \ac{ANN} algorithm for the K-d tree used in this work is in Algorithm \ref{alg:getNeighbors}. The algorithm computes the distance between the data sample $q$ and the current search node only if the node is active (i.e., not flagged as deleted), as in Lines 3-4. If the neighbors' list has fewer than $k$ instances, the instance of the current node is added to the neighbors list in Line 6. Otherwise, the current maximum distance of the neighbors' list is compared to the distance of the current node's instance in Line 9. If smaller, it replaces the farthest node in Line 10. The procedure to choose the next subtree to search works as follows: In line 11, the value of $q$ in the split dimension $s$ is compared to $I[s]$. If it is equal or greater, the best node to continue the search is the right one (Line 12). Otherwise, it will be the left child (Line 15). The split criterion is tested in Line 18: if the current maximum distance in the neighbors' list is less than $c$, the subtree is searched for neighbors. On the contrary, it is pruned. For the Minkowski-based distance functions (Euclidean and Manhattan), the adopted split criterion is $c=|\mathbf{q}[s]-\mathbf{I_t}[s]|$. Backtracking is used to check other subtrees for potential neighbors that satisfy the split criterion (Lines 21-22). Implementation details to deal with leaf nodes were omitted for simplicity. The source code is available in our GitHub repository\footnote{Omitted for anonymity.}.

\begin{algorithm}[htb]
\footnotesize
\SetKwInOut{Input}{Input}\SetKwInOut{Output}{Output}

\Input{Node, query $\mathbf{q}$, neighbors, $k$}
\Output{Updated neighbors}

\If{Node is null}{\Return neighbors}

\If{Node is active}{
    UpdateNeighbors(neighbors, Node.instance, $\mathbf{q}$, $k$)
}

$s \gets Node.splitDimension$\\
\If{$\mathbf{q}[s] \geq Node.instance[s]$}{
    $(best, other) \gets (Node.right, Node.left)$
}\Else{
    $(best, other) \gets (Node.left, Node.right)$
}

$neighbors \gets ANN(best, \mathbf{q}, neighbors, k)$\\

\If{ShouldSearchSubtree(Node, $\mathbf{q}$, s, neighbors)}{
    $neighbors \gets ANN(other, \mathbf{q}, neighbors, k)$
}

\Return $neighbors$

\caption{ANN}
\label{alg:getNeighbors}
\end{algorithm}

\noindent\textbf{Adapting to the Canberra Distance.} We adapt the K-d tree to the Canberra distance because \cite{barboza2023} demonstrated that it usually leads to better results for data streams. Thus, a K-d tree for data streams that supports the Canberra distance offers benefits for data stream applications. First, consider the Canberra distance function in Equation \ref{eq:canberra}.

\begin{equation}
\label{eq:canberra}
    d(\mathbf{x_1},\mathbf{x_2})=\sum^{n}_{i=1}\frac{|\mathbf{x_1}[i]-\mathbf{x_2}[i]|}{|\mathbf{x_1}[i]|+|\mathbf{x_2}[i]|}
\end{equation}

For using it to \ac{ANN}, we can use the same idea of the Minkowski-based K-d tree, and adopt a dimension-wise split heuristic. Considering the split dimension $s$ of a node $t$, the split criterion $c$ of the K-d tree with the Canberra distance we use is $c = \frac{|\mathbf{q}[s] - \mathbf{I_t}[s]|}{|\mathbf{q}[s]| + |\mathbf{I_t}[s]|}$.

% \begin{equation}
% \label{eq:split-crit-canberra}
%     c = \frac{|\mathbf{q}[s] - \mathbf{I_t}[s]|}{|\mathbf{q}[s]| + |\mathbf{I_t}[s]|}
% \end{equation}

The drawback of this strategy lies in the curse of dimensionality. When applied to higher-dimensional datasets, a single dimension is unlikely to be sufficient to effectively prune subtrees from the search space.
% To mitigate this, an approximation parameter $\epsilon$ was applied to the equation, which can be adjusted to offer a looser or tighter threshold when pruning subtrees.

% The $\epsilon$ parameter was included in the equation to dictate how hard the split of subtrees will be. Since we use a single dimension to prune subtrees, the search space may not be properly decreased in high-dimensional datasets. Further, if one wants to have a tighter or looser pruning of subtrees, the $\epsilon$ parameter can be adjusted depending on the dataset or application.

% \begin{table}[htb]
%     \centering
%     \caption{Distance Functions and its split criterion.}
%     \label{tab:distance-functions}
%     \begin{tabular}{c c c}
%     \toprule
%         Distance Function & Equation & Split Criterion \\
%         \midrule
%         Euclidean & $\displaystyle d(x,y)=\sqrt{\sum_{i=1}^{n}(x_i-y_i)^2} $ & $D\leq|\mathbf{t}_s-\mathbf{x}_s|$ \\
%         Manhattan & $\displaystyle d(x,y)=\sum_{i=1}^{n}|x_i-y_i|$ & $D\leq|\mathbf{t}_s-\mathbf{x}_s|$ \\ 
%         Canberra & $\displaystyle d(x,y)=\sum^{n}_{i=1}\frac{|x_i-y_i|}{|x_i|+|y_i|}$ \\
%         \bottomrule
%     \end{tabular}
% \end{table}

%% file: experiments.tex
\section{Experiments}
\label{sec:experiments}

The experiments in this paper were designed to answer the following Research Questions:

\begin{itemize}
    \item \textbf{RQ1:} How does the K-d tree behave in different data stream scenarios?
    \item \textbf{RQ2:} How do crucial dynamic operations in data streams (insertion/deletion) impact the K-d tree?
    \item \textbf{RQ3:} What are the drawbacks and challenges of using K-d tree in data streams?
\end{itemize}

We evaluated the accuracy of the online K-d tree algorithm against brute-force kNN for each distance function. Additionally, we measure the average precision of the neighbors computed by the K-d tree. Finally, we report the average number of instances processed per second of each algorithm.

All algorithms were initialized with 5,000 instances and used a sliding window of up to 20,000 instances, updated in a test-then-train manner \cite{gama2014}. In this work, the $k$ value was set to $k=11$ without any optimization, as the focus is on comparing the brute-force \ac{kNN} to the K-d tree rather than fine-tuning $k$ for performance. The $\gamma$ was set to $\gamma=2$, i.e., the K-d tree is rebuilt when it doubles in size, a strategy also adopted by \cite{muja2009}, and the $\beta$ hyperparameter was set to $\beta=0.3$ (if 30\% of the nodes are deleted, rebuild the K-d tree), as in \cite{barboza2025}.

We used a combination of synthetic and real-world datasets widely used in the data stream literature, with varying sample size and dimensionality. We evaluated on 6 real-world datasets (Electricity \cite{moa}, NOAA \cite{ditzlerEtAl2012}, Insects-AB \cite{Souza2020}, Covertype \cite{moa}, Gas Sensor \cite{uci}, Dry Bean\cite{uci}) ranging from 8 to 128 features, and 3 synthetic datasets (SEA \cite{street2001}, Sine \cite{gama2004}, RandomRBF \cite{bifet2009a}) with induced concept drift. More information about the datasets can be found in the repository.
% \footnote{https://github.com/eduardovlb/OKDTree}

The synthetic datasets were generated using the \ac{MOA} framework \cite{moa}. Concept drift was simulated on SEA and Sine by changing the generating function every 25,000 instances. Virtual drift was induced on Dry Bean following \cite{almeidaEtAl2018}. All experiments were conducted on a machine running Ubuntu 24.04 with an Intel Core i7 processor and 16GB of DDR4 RAM. The experiments were executed using Java 17 on the \ac{MOA} framework \cite{moa}. 
No multi-threading was implemented during the experiments. 

\noindent\textbf{The impact of dimensionality.} We compare the accuracy and neighborhood precision of the Online K-d tree against brute-force \ac{kNN} under different distance functions. Table~\ref{tab:acc-kdtree} reports classification accuracy, while Table~\ref{tab:prec-kdtree} reports neighborhood precision, defined as the percentage of neighbors identical to those returned by brute-force \ac{kNN}. A precision of 100\% indicates identical neighborhoods.

Overall, the Online K-d tree achieves slightly lower average accuracy than brute-force \ac{kNN}, as expected from approximate neighborhood search. However, differences in accuracy are generally small and highly dataset- and distance-function-dependent. In particular, the Canberra distance achieves the best average accuracy and neighborhood precision across datasets, performing especially well on Electricity and Insects-AB. This highlights the importance of supporting multiple distance functions, as the optimal choice varies by domain.

\begin{table*}[htb]
    \centering
    % \scriptsize
    \caption{Accuracies of kNN and Online K-d tree with different distance functions. Values in bold are the best results for each distance function.}
    \label{tab:acc-kdtree}
    \begin{tabular}{l c c | c c | c c}
    \toprule
         & \multicolumn{2}{c|}{Euclidean} & \multicolumn{2}{c|}{Manhattan} & \multicolumn{2}{c}{Canberra} \\
        Dataset & kNN & OK-d Tree & kNN & OK-d Tree & kNN & OK-d Tree \\
        \midrule
        NOAA & \textbf{79.52} & 78.68 & \textbf{79.50} & 79.45 & \textbf{77.08} & 76.91 \\
        Electricity & 75.26 & \textbf{75.57} & \textbf{76.70} & 76.48 & \textbf{81.31} & 80.51 \\
        Insects-AB & \textbf{56.54} & \textbf{56.54} & 57.73 & \textbf{58.02} & 58.87 & \textbf{59.17} \\
        Covertype & \textbf{90.29} & 87.39 & \textbf{90.80} & 89.02 & \textbf{90.28} & 89.42 \\
        Gas Sensor & \textbf{94.02} & 91.57 & \textbf{94.40} & 93.71 & \textbf{97.95} & \textbf{97.95} \\
        Dry Bean & 68.00 & \textbf{74.92} & 74.57 & \textbf{76.02} & \textbf{90.15} & 90.13 \\
        SEA & \textbf{85.69} & 85.13 & \textbf{85.67} & 85.25 & \textbf{85.20} & 84.96 \\
        Sine & \textbf{71.32} & 68.59 & \textbf{71.37} & 69.32 & \textbf{70.97} & 68.70 \\
        RandomRBF & \textbf{95.72} & 95.56 & \textbf{95.26} & \textbf{95.26} & \textbf{93.74} & \textbf{93.74} \\\hline
        Average & \textbf{82.10} & 81.76 & \textbf{82.96} & 82.59 & \textbf{84.63} & 84.22 \\
        \bottomrule
    \end{tabular}
\end{table*}

\begin{table}[ht]
    \centering
    \caption{Neighborhood precision (\%) of Online K-d tree compared to kNN on different distance functions.}
    \label{tab:prec-kdtree}
    \begin{tabular}{l c c c c c c c c c c}
    \toprule
    Distance & NOAA & Elec. & Insects & Covertype & GasS. & DryB. & SEA & Sine & RBF & Average \\
    \midrule
    Euclidean & 64.01 & 41.22 & 85.06 & 63.41 & 71.80 & 31.23 & 65.30 & 62.23 & 72.06 & 61.81 \\
    Manhattan & \textbf{96.57} & 49.73 & 95.69 & 78.00 & 87.78 & 38.61 & \textbf{74.32} & \textbf{77.35} & \textbf{88.74} & 76.33 \\
    Canberra & 83.23 & \textbf{78.65} & \textbf{96.51} & \textbf{85.56} & \textbf{100} & \textbf{98.22} & 62.73 & 63.52 & 88.40 & \textbf{84.09} \\
    % \midrule
    % Average & 61.81 & 76.33 & \textbf{84.09} &  &  &  &  &  &  \\
    \bottomrule
    \end{tabular}
\end{table}

On the Gas Sensor dataset, Euclidean distance shows a larger accuracy gap, likely due to ineffective pruning caused by small feature magnitudes in high-dimensional space. Manhattan distance mitigates this issue, while Canberra achieves the highest accuracy but also a neighborhood precision of 100\%, indicating limited pruning.

On the Dry Bean dataset, the Online K-d tree outperforms brute-force \ac{kNN} under Euclidean and Manhattan distances despite low neighborhood precision. This suggests that axis-aligned partitions restrict the search to more locally relevant regions, filtering less informative points. The large variation in feature magnitudes may amplify this effect, whereas Canberra is less affected due to its normalization factor.

For synthetic datasets with concept drift, accuracy differences are negligible on SEA but more pronounced on Sine. This may be attributed to the insertion of new-concept instances into deeper tree regions, which are harder to reach for approximate search. This effect appears stronger when concept boundaries are sharper, as in Sine, reinforcing the need for drift-aware adaptation mechanisms.

High-dimensional datasets seem to struggle in non-optimal neighborhoods. To overcome this, an approximation parameter $\epsilon$ can be included in the split criterion equation, aiming to adjust the rigor of the pruning of subtrees, or one can define a minimum number of visited nodes according to the search space $n$, a strategy previously adopted by \cite{muja2009}.

\noindent\textbf{Instances per second.} Table~\ref{tab:is-kdtree} reports the average number of instances processed per second. Overall, the Online K-d tree significantly outperforms brute-force \ac{kNN} in instances per second, achieving speedups of up to 7.7x, 7.2x, and 6.4x on average for Euclidean, Manhattan, and Canberra distances, respectively. 

Performance gains depend on both dimensionality and data distribution. The K-d tree consistently excels on low-dimensional datasets such as SEA and Sine, while results on higher-dimensional datasets vary. For example, despite its dimensionality, Covertype benefits from effective pruning, despite non-optimal neighborhoods leading to drops in accuracy, whereas datasets such as RandomRBF and Insects-AB show limited gains. These results indicate that pruning efficiency depends not only on dimensionality but also on the spatial structure of the data.

% Now, let us evaluate how the number of instances processed per second varied between the brute force \ac{kNN} and the K-d tree. The results are exposed in Table \ref{tab:is-kdtree}. We see that, on average, the Online K-d tree (OK-d Tree in the Table) was able to process more instances than \ac{kNN} in one second. With the Euclidean and Manhattan distances, the K-d tree processed over seven times more instances than \ac{kNN}, and over 6 times with the Canberra distance.

\begin{table*}[htb]
    \centering
    % \footnotesize
    % \renewcommand{\arraystretch}{0.85}
    % \setlength{\aboverulesep}{0pt}
    % \setlength{\belowrulesep}{0pt}
    \caption{Average number of instances processed per second (I/s) of kNN and Online K-d tree with different distance functions. Values in bold are the best results for each distance function. Reported results are an average of 10 runs.}
    \label{tab:is-kdtree}
    \begin{tabular}{l c c c | c c c | c c c}
    \toprule
         & \multicolumn{3}{c|}{Euclidean} & \multicolumn{3}{c|}{Manhattan} & \multicolumn{3}{c}{Canberra} \\
        Dataset & kNN & OK-d Tree & Gain & kNN & OK-d Tree & Gain & kNN & OK-d Tree & Gain \\
        \midrule
        NOAA & 1,713 & \textbf{3,490} & 2x & \textbf{1,597} & 801 & 0.5x & 1,352 & \textbf{2,082} & 1.5x \\
        Electricity & 777 & \textbf{8,210} & 10.6x & 766 & \textbf{5,256} & 6.9x & 644 & \textbf{2,833} & 4.4x \\
        Insects-AB & 292 & \textbf{348} & 1.2x & \textbf{280} & 89 & 0.3x & \textbf{263} & 110 & 0.4x \\
        Covertype & 187 & \textbf{454} & 2.4x & 179 & \textbf{219} & 1.2x & 165 & \textbf{291} & 1.8x \\
        Gas Sensor & 207 & \textbf{212} & 1.0x & 193 & \textbf{293} & 1.5x & \textbf{158} & 86 & 0.5x \\
        Dry Bean & 1,345 & \textbf{5,382} & 4.0x & 1,319 & \textbf{6,285} & 4.8x & 1,115 & \textbf{2,861} & 2.6x \\
        SEA & 2,082 & \textbf{33,333} & 16.0x & 2,078 & \textbf{34,926} & 16.8x & 1,875 & \textbf{30,744} & 16.4x \\
        Sine & 1,892 & \textbf{23,342} & 12.3x & 1,911 & \textbf{21,839} & 11.4x & 1,681 & \textbf{13,808} & 8.2x \\
        RandomRBF & 1,411 & \textbf{1,642} & 1.2x & \textbf{1,392} & 569 & 0.4x & \textbf{1,112} & 489 & 0.4x \\ \hline
        Average & 1,101 & \textbf{8,490} & 7.7x & 1,079 & \textbf{7,809} & 7.2x & 929 & \textbf{5,923} & 6.4x \\
        \bottomrule
    \end{tabular}
\end{table*}

\noindent\textbf{Lessons Learned.} The experiments highlight three main lessons. First, the K-d tree's effectiveness is highly data-dependent: it efficiently reduces the search space on low-dimensional datasets, while benefits diminish on some high-dimensional ones. Dimensionality alone does not fully explain this; the data distribution also influences subtree pruning \textbf{(RQ1)}.

Second, dynamic operations such as insertion and deletion can be performed in $O(\log n)$ when the structure remains balanced. The adopted rebuild heuristic helped maintain balance throughout the experiments. Since the evaluation followed a test-then-train protocol, insertions and deletions were frequent; in real-world streams with delayed labels, these operations are typically less common, further favoring the online K-d tree \textbf{(RQ2)}.

Third, the main challenges of using K-d trees in data streams relate to balancing, deletions, and concept drift. Periodic rebuilding mitigates imbalance at the cost of limited overhead, while lazy deletion avoids costly subtree reinsertion. Memory cleanup was handled by rebuilding the tree after 30\% of the nodes were marked deleted. Under concept drift, older concepts tend to populate deeper regions of the tree and are more likely to be pruned, indicating the need for adaptive mechanisms in highly non-stationary streams \textbf{(RQ3)}.

Overall, the Online K-d tree reduces processing time in many streaming scenarios, but its use should be guided by data characteristics and neighborhood quality. Supporting the Canberra distance extends applicability to domains where alternative distance functions improve performance.

%% file: conclusion.tex
\section{Conclusion}
\label{sec:conclusion}

This work described an online K-d tree algorithm for data streams that supports dynamic operations through incremental updates, lazy deletion, and periodic rebuilding. We also described the adaptation of the approximate neighborhood search to support the Canberra distance, enabling its use in domains where this metric yields better neighborhoods. Experimental results demonstrate that the online K-d tree significantly reduces processing time compared to brute-force \ac{kNN}, while incurring only negligible accuracy loss on most datasets, making it a practical alternative for large-scale streaming applications. 

Future work includes extending the approach to additional distance functions and exploring complementary strategies, such as dimensionality reduction, to further improve neighborhood search efficiency under data streams.

\section*{Acknowledgments}

This research has been supported by NSRC-Canada (Natural Science and Engineering Research Council of Canada) grant RGPIN-2022-02943.

% This work has presented an online K-d tree algorithm that can perform real-time dynamic operations. It avoids potential issues found under data streams caused by changing data distribution, which may lead to an unbalanced structure, and the deletion of nodes, which would require additional complexity to deal with replaced nodes, which could violate the K-d tree's invariants. We also have adapted the \ac{ANN} algorithm of the K-d tree to work with the Canberra distance function, opening space for the K-d tree to be applied in more domains where the Canberra distance leads to better neighborhoods.

% The experiments performed in this work showed that the online K-d tree had an average smaller processing time with a negligible loss in accuracy. Thus, we hope large-scale applications will benefit from this work's contributions. In the future, we intend to explore more data structures and options to optimize the processing time of the neighborhood search under data streams and to explore adaptations of the K-d tree to more distance functions. Additionally, dimensionality reduction methods may also improve results regarding neighborhood search in data streams and are a prospect for future work.